\documentclass[runningheads]{llncs}

\usepackage{url}
\usepackage{amsfonts,amsmath} 
\usepackage{xspace}
\usepackage{paralist}

\usepackage{xcolor}         

\usepackage{algorithm}
\usepackage{algorithmic}
\usepackage{booktabs}
\usepackage[english]{babel}
\usepackage{cleveref}

\usepackage{tikz}
\usepackage{tkz-graph}
\usetikzlibrary{
    arrows,
    arrows.meta,
    fadings,
    patterns,
    positioning,
    shapes.geometric
}

\usetikzlibrary{arrows,topaths,calc}
\usetikzlibrary{positioning}
\usetikzlibrary{shadows}
\usetikzlibrary{shapes}
\usetikzlibrary{backgrounds}
\usetikzlibrary{decorations.pathreplacing}

\newcommand{\cellwidth}{0.46\textwidth}
\definecolor{babyblue}{RGB}{53,172,229}
\definecolor{turquoise}{RGB}{0, 245, 255}

\usepackage[nolist]{acronym}
\begin{acronym}[UML]
	\acro{AI}{Artificial Intelligence}
	\acro{MTL}{Multi-task Learning}
	\acro{CEL}{Class Expression Learning}
	\acro{CL}{Concept Learning}
	\acro{DL}{Description Logic}
    \acro{KB}{Knowledge Base}
    \acro{KG}{Knowledge Graph}
    \acrodefplural{KG}{Knowledge Graphs}
    \acro{KGE}{Knowledge Graph Embedding}
	\acro{ILP}{Inductive Logic Programming}
	\acro{RL}{Reinforcement Learning}
	\acro{OWL}{Web Ontology Language}
	\acro{SW}{Semantic Web}
	\acro{OWA}{Open World Assumption}
	\acro{CWA}{Close World Assumption}
\end{acronym}  

\newcommand{\ABox}{\ensuremath{Abox}\xspace}
\newcommand{\TBox}{\ensuremath{Tbox}\xspace}
\newcommand{\kb}{\ensuremath{\mathcal{K}}\xspace}

\newcommand{\NamedConcepts}{\ensuremath{N_C}\xspace}
\newcommand{\Individuals}{\ensuremath{N_I}\xspace} 
\newcommand{\Roles}{\ensuremath{N_R}\xspace}

\newcommand{\Nero}{\textsc{NeRo}\xspace}
\newcommand{\targetexpressions}{\ensuremath{\mathcal{T}}\xspace}
\newcommand{\Retrievalfunc}{\ensuremath{\mathcal{R}}}
\newcommand{\ALC}{\ensuremath{\mathcal{ALC}}\xspace}

\newcommand{\AllALCConcepts}{\ensuremath{\mathcal{C}}\xspace}

\newcommand{\heuristic}{\ensuremath{\phi}\xspace}

\newcommand{\StateSpace}{\ensuremath{\mathcal{S}}\xspace}

\usepackage{graphicx}
\begin{document}
\title{Learning Permutation-Invariant Embeddings \\for Description Logic Concepts}

\author{Caglar Demir \and Axel-Cyrille Ngonga Ngomo}
\institute{Data Science Research Group, Paderborn University}

\maketitle
\begin{abstract}
Concept learning deals with learning description logic concepts from a background knowledge and input examples. 
The goal is to learn a concept that covers all positive examples, while not covering any negative examples.
This non-trivial task is often formulated as a search problem within an infinite quasi-ordered concept space.
Although state-of-the-art models have been successfully applied to tackle this problem, their large-scale applications have been severely hindered due to their excessive exploration incurring impractical runtimes.
Here, we propose a remedy for this limitation.
We reformulate the learning problem as a multi-label classification problem and propose a neural embedding model (NERO) that learns permutation-invariant embeddings 
for sets of examples tailored towards
predicting $F_1$ scores of pre-selected description logic concepts.
By ranking such concepts in descending order of predicted scores, a possible goal concept can be detected within few retrieval operations, i.e., no excessive exploration.
Importantly, top-ranked concepts can be used to
start the search procedure of state-of-the-art symbolic models in multiple advantageous regions of a concept space, rather than starting it in the most general concept $\top$.
Our experiments on 5 benchmark datasets with 770 learning problems firmly suggest that NERO significantly (p-value $<1\%$) outperforms the state-of-the-art models in terms of $F_1$ score, the number of explored concepts, and the total runtime.
We provide an open-source implementation of our approach.\footnote{\raggedright\url{https://github.com/dice-group/Nero}}
\end{abstract}
\section{Introduction}
\label{sec:introduction}
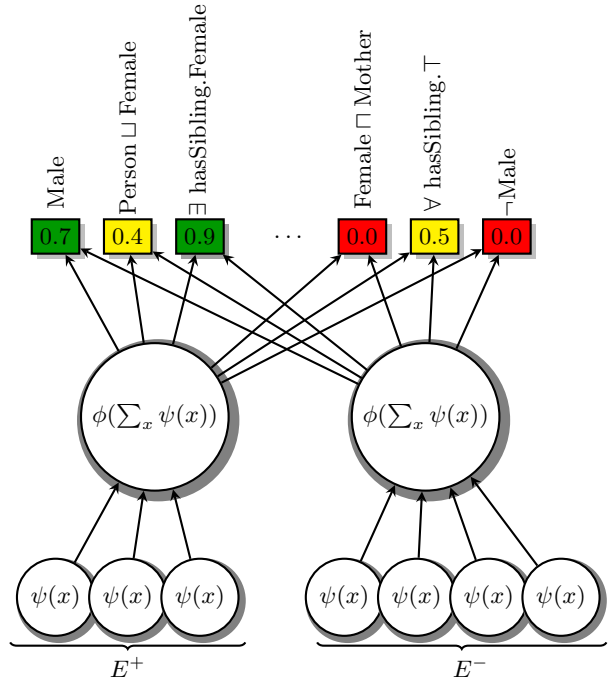
\begin{figure}
    \centering
    \begin{tikzpicture}[scale=1.2,baseline,thick]
  \GraphInit[vstyle=Classic]
    \tikzset{vertex/.style={draw=black,shape=circle,fill=white,minimum size=13pt,circular drop shadow}}
    \tikzset{box/.style={draw=black,shape=rectangle,fill=white,minimum size=13pt,drop shadow}}
  \node at (.2, 0)[vertex] (eplus1) {$\psi(x)$};
  \node at (1., 0)[vertex] (eplus2) {$\psi(x)$};
  \node at (1.8, 0)[vertex] (eplus3) {$\psi(x)$};
  \node at (-0.2, -0.5) (u1l) {};
  \node at (2.2, -0.5) (u2l) {};
  \node at (1.0, -0.775) (subl) {$E^+$};
  \draw[decoration={brace,mirror},decorate] (u1l.west) -- (u2l.east);

  \node at (3.4,0)[vertex] (eminus1) {$\psi(x)$};
  \node at (4.2,0)[vertex] (eminus2) {$\psi(x)$};
  \node at (5.0,0)[vertex] (eminus3) {$\psi(x)$};
  \node at (5.8,0)[vertex] (eminus4) {$\psi(x)$};

  \node at (3.2,-0.5) (v1l) {};
  \node at (6.2,-0.5) (v2l) {};
  \node at (4.8,-0.775) (subl) {$E^-$};
  \draw[decoration={brace,mirror},decorate] (v1l.west) -- (v2l.east);

  \node at (1.3,2.)[vertex] (b1) {$\phi (\sum_x \psi(x))$};
  \node at (4.3,2.)[vertex] (b2) {$\phi (\sum_x \psi(x))$};

  \node at (.2,4.0)[box,fill=black!40!green] (out1) {0.7};
   \node[rotate=90] at (0.2,4.6) (out1name) {Male};
  
  \node at (1.,4.0)[box,fill=yellow] (out2) {0.4};
   \node[rotate=90] at (1.,5.2) (out1malename2) {$\text{Person}\sqcup\text{Female}$};

  \node at (1.8,4.0)[box,box,fill=black!40!green] (out1male) {0.9};
   \node[rotate=90] at (1.8,5.4) (out1malename) {$\exists\text{ hasSibling.Female}$};

  \node at (2.8,4.0) (outmiddle) {\dots};
  \node at (3.6,4.0)[box,fill=red] (out4) {0.0};
   \node[rotate=90] at (3.6,5.3) (out11name) {$\text{Female}\sqcap \text{Mother}$};
  \node at (4.4,4.0)[box,fill=yellow] (out5) {0.5};
  \node[rotate=90] at (4.4,5.15) (out1name) {$\forall\text{ hasSibling.}\top$};

  \node at (5.2,4.0)[box,fill=red] (out6) {0.0};
   \node[rotate=90] at (5.2,4.6) (out1name) {$\neg \text{Male}$};

  
  \tikzstyle{EdgeStyle}=[->,>=stealth,thick]
  \Edge (eplus1)(b1) 
  \Edge (eplus2)(b1) 
  \Edge (eplus3)(b1) 
 
  \Edge (eminus1)(b2)
  \Edge (eminus2)(b2) 
  \Edge (eminus3)(b2) 
  \Edge (eminus4)(b2)

  \Edge (b1)(out1) \Edge (b1)(out2) \Edge (b1)(out1male) \Edge (b1)(out4)\Edge (b1)(out5)\Edge (b1)(out6) 

  \Edge (b2)(out1) \Edge (b2)(out2) \Edge (b2)(out1male) \Edge (b2)(out4)\Edge (b2)(out5)\Edge (b2)(out6) 

\end{tikzpicture}

    \caption{Visualization of \Nero.
    Boxes and values denote the pre-selected unique \ac{DL} concepts and their predicted $F_1$ scores, respectively.}
    \label{fig:nero}
\end{figure}

Deep learning based models have been effectively applied to tackle various graph-related problems, including question answering, link prediction~\cite{hogan2021knowledge,nickel2015review}.
Yet, their predictions are not human-interpretable and confined within a fixed set vocabulary terms~\cite{demir2021convolutional,dettmers2018convolutional}. 
In contrast, \acp{DL} provide means to derive human-interpretable inference in an infinite setting~\cite{baader2003description,Heindorf2021EvoLearner,lehmann2010concept}.
Deriving explanations for \acp{DL} concepts has been long understood~\cite{borgida2000explaining}.
For instance, explanations can be derived by using the subsumption hierarchy as a sequence of binary classifiers in a fashion akin to following a path in decision tree~\cite{westphal2021simulated,bin2017implementing}.
Utilizing \acp{DL} is considered as a possible backbone for explainable \ac{AI}~\cite{ijcai2021p281}.
Although \acp{DL} have become standard techniques to formalize \ac{KB}~\cite{hogan2021knowledge,horrocks2003shiq,michel2019efficient},
the highly incomplete nature of \acp{KB} and impractical runtimes of symbolic models have been a challenge for fulfilling its potential.
State-of-the-art \ac{CL} models have been successfully applied to learn \ac{DL} concepts from a \ac{KB} and input examples~\cite{lehmann2009dl,lehmann2010concept}.
Yet, their practical applications have been severely hindered by their impractical runtimes.
This limitation stems from the reliance of myopic heuristic function that often incurs excessive exploration of concepts~\cite{Heindorf2021EvoLearner,kouagou2021learning,westphal2021simulated}.
A \ac{DL} concept is explored by retrieving its individuals and calculating its quality w.r.t. input \ac{KB} and examples (see~\Cref{sec:background}).
As the size of an input \ac{KB} grows, excessive exploration has been a computational bottleneck in practical applications.
Here, we propose a remedy for this limitation.
We reformulate the learning problem as a multi-label classification problem and propose \Nero--a \underline{ne}u\underline{r}al permutati\underline{o}n-invariant embedding model.
Given a set of positive examples $E^+$ and a set of negative examples $E^-$,~\Nero predicts $F_1$ scores of pre-selected \ac{DL} concepts as shown in~\Cref{fig:nero}.

By ranking pre-selected \ac{DL} concepts in descending order of predicted scores, a goal concept can be found by only exploring few top-ranked concepts.
Importantly, top-ranked concepts can be used to initialize the standard search procedure of state-of-the-art models, if a goal concept is not found.
By this, a state-of-the-art \ac{CL} model is endowed with the capability of starting the search in more advantageous states, instead of starting it in the most general concept $\top$.
Our experiments on 5 benchmark datasets with 770 learning problems indicate that \Nero \textbf{significantly} (p-value $<1\%$) outperforms the state-of-the-art models in standard metrics such as $F_1$ score, the number of explored concepts, and the total runtime.
Importantly, equipping \Nero with a state-of-the-art model (CELOE) further improves $F_1$ scores on benchmark datasets with a low runtime cost.
The results of Wilcoxon signed rank tests confirm that the superior performance of~\Nero is significant.
We provide an open-source implementation of~\Nero, including pre-trained models, evaluation scripts as well as a web service.\footnote{\url{https://github.com/dice-group/Nero}}
\section{Background}
\label{sec:background}
\subsubsection{Knowledge Base:} 
\begin{figure}
    \centering
    \small
\tikzset{
    man/.pic={    
        \node[babyblue, circle, fill, minimum size = 1.5mm, inner sep = 0] (head) at (0,0) {};
        \node[
            babyblue,
            draw,
            fill,
            rectangle,
            rounded corners = 0.2mm,
            minimum width = 1.2mm,
            minimum height = 2.4mm,
            below = 0.1mm of head,
            inner sep = 1pt
        ] (body) {};
        \draw[babyblue, line width = 0.5mm,  round cap-round cap]
            ([shift={(-0.42mm, 0.45mm)}]body.south) --++(-90:2.4mm);
        \draw[babyblue, line width = 0.5mm,round cap-round cap]
            ([shift={(0.42mm, 0.45mm)}]body.south) --++(-90:2.4mm);
        \draw[babyblue, line width = 0.2mm, round cap-round cap, rounded corners]
            ([shift = {(-0.85mm, -0.2mm)}]body.north) -- ++(0mm, -2mm);
        \draw[babyblue, line width = 0.2mm, round cap-round cap, rounded corners]
            ([shift = {(0.85mm, -0.2mm)}]body.north) -- ++(0mm, -2mm);
    }
}
\tikzset{
    woman/.pic={    
        \node[magenta, circle, fill, minimum size = 1.5mm, inner sep = 0] (head) at (0,0) {};
        \node[
            magenta,
            transform shape,
            draw,
            fill,
            trapezium,
            trapezium angle=65,
            trapezium stretches=true,
            rounded corners = 0.2mm,
            minimum width = 2mm,
            minimum height = 2.8mm,
            below = 0.1mm of head,
            inner sep = 1pt
        ] (body) {};
        \draw[magenta, line width = 0.5mm,  round cap-round cap]
            ([shift={(-0.35mm, 0.45mm)}]body.south) --++(-90:2mm);
        \draw[magenta, line width = 0.5mm,round cap-round cap]
            ([shift={(0.35mm, 0.45mm)}]body.south) --++(-90:2mm);
        \draw[magenta, line width = 0.2mm, round cap-round cap, rounded corners]
            ([shift = {(-0.8mm, -0.2mm)}]body.north) -- ++(-0.35mm, -2mm);
        \draw[magenta, line width = 0.2mm, round cap-round cap, rounded corners]
            ([shift = {(0.8mm, -0.2mm)}]body.north) -- ++(0.35mm, -2mm);
    }
}
\newcommand{\Male}[4]{%
    \node[label = {[font = \sf\tiny, label distance = -1mm]below:#4}] (#1) at (#2, #3) {%
        \begin{tikzpicture}
            \pic at (0, 0) {man};
        \end{tikzpicture}
    };
}
\newcommand{\Female}[4]{%
    \node[label = {[font = \sf\tiny, label distance = -1mm]below:#4}] (#1) at (#2, #3) {%
        \begin{tikzpicture}
            \pic at (0, 0) {woman};
        \end{tikzpicture}
    };
}


\begin{tikzpicture}[
                    xscale = 1.4, 
                    yscale = 1.8, 
                    TreeLine/.style = {
                        line width = 0.4mm,
                        black!70
                    },
                    HighlightLine/.style = {
                        TreeLine,
                        orange
                    },
                    InferredLine/.style = {
                        HighlightLine,
                        dashed
                    },
                    Marriage/.style = {
                        circle,
                        fill = black!70,
                        inner sep = 0pt,
                        minimum width = 3
                    }
                ]
                
                    \Male  {idF10M171}{1}{3}{F10M171}
                    \Female{idF10F172}{2}{3}{F10F172}
                    
                    \Male  {idF10M180}{0}{2}{F10M180}
                    \Female{idF10F179}{1}{2}{F10F179}
                    \Male  {idF10M173}{2}{2}{F10M173}
                    \Female{idF10F174}{3}{2}{F10F174}

                    \Female{idF10F177}{2}{1}{F10F177}
                    \Female{idF10F175}{3}{1}{F10F175}

                    \draw[TreeLine] (idF10M171) -- (idF10F172);
                    \draw[TreeLine] (idF10M180)   -- (idF10F179);
                    \draw[TreeLine] (idF10M173)  -- (idF10F174);    
        
                    \node[Marriage] at (1.5, 3) {};            
                    \draw[TreeLine] (1.5, 2.6) -- (1.5, 3);    
                    \draw[TreeLine] (idF10F179)                   
                                    -- (1, 2.4)
                                    arc (180:90:0.1)
                                    -- (1.4, 2.5)
                                    arc (270:360:0.1)
                                    arc (180:270:0.1)
                                    -- (1.9, 2.5)
                                    arc (90:0:0.1)
                                    -- (idF10M173);

                    \node[Marriage] at (0.5, 2) {};            
                    
                    \node[Marriage] at (2.5, 2) {};            
                    \draw[TreeLine] (2.5, 1.6) -- (2.5, 2);    
                    \draw[TreeLine] (idF10F177)                   
                                    -- (2, 1.4)
                                    arc (180:90:0.1)
                                    -- (2.4, 1.5)
                                    arc (270:360:0.1)
                                    arc (180:270:0.1)
                                    -- (2.9, 1.5)
                                    arc (90:0:0.1)
                                    -- (idF10F175);

                    \node[Marriage, black] at (1.5, 3) {};  
                    \node[Marriage, black] at (2.5, 2) {};  
                    \draw[TreeLine] (idF10F179)
                                         -- (1, 2.4)
                                         arc (180:90:0.1)
                                         -- (1.4, 2.5)
                                         arc (270:360:0.1)
                                         -- (1.5, 3)
                                         -- (idF10F172)
                                         -- (1.5, 3)
                                         -- (1.5, 2.6)
                                         arc (180:270:0.1)
                                         -- (1.9, 2.5)
                                         arc (90:0:0.1)
                                         -- (idF10M173)
                                         -- (2.5, 2)
                                         -- (2.5, 1.6)
                                         arc (180:270:0.1)
                                         -- (2.9, 1.5)
                                         arc (90:0:0.1)
                                         -- (idF10F175);

                \end{tikzpicture}
        \tikzset{BaseCell/.style = {
                draw = white,
                inner sep = 5pt,
                font = \tiny\sf,
                align = left,
                line width = 1pt,
                rounded corners = 3pt
            }}
        \tikzset{
            Example/.style = {
                BaseCell,
                draw = white,
                fill = black!70,
                text = white
            }
        }%
        \tikzset{
            Graph/.style = {
                fill = black!20
            }
        }%
        \tikzset{
            KB/.style = {
                BaseCell,
                fill = white
            }
        }%
        \begin{tikzpicture}
        [Line/.style = {inner sep = 2pt,line width = 0pt}]        
            \node[Line] (line1) {%
                \begin{tikzpicture}
                    \node[KB] (kb1) {%
                        \begin{minipage}{\cellwidth}
                            \textbf{TBox:} \\
                            Brother $\sqsubseteq$Male\\
                            Brother $\sqsubseteq$ 
                            PersonWithASibling\\
                            Child $\sqsubseteq$ Person\\
                            Daughter $\sqsubseteq$ Child, Daughter $\sqsubseteq$ Female\\
                            Father $\sqsubseteq$ Male, Father $\sqsubseteq$ Parent\\
                            Female $\sqsubseteq$ Person\\
                            Grandchild $\sqsubseteq$ Child\\
                            Granddaughter $\sqsubseteq$ Female\\ Granddaughter $\sqsubseteq$ Grandchild\\
                            Grandfather$\sqsubseteq$ Grandparent\\ Grandfather$\sqsubseteq$ Male\\
                            Grandmother$\sqsubseteq$Female\\ 
                            Grandmother$\sqsubseteq$ Grandparent\\
                            Grandparent$\sqsubseteq$Parent\\
                            Grandson$\sqsubseteq$Grandchild,Grandson$\sqsubseteq$Male\\
                            Male $\sqsubseteq$ Person\\
                            Mother $\sqsubseteq$ Person, Mother $\sqsubseteq$ Parent\\
                            Parent $\sqsubseteq$ Person\\                            PersonWithASibling $\sqsubseteq$ Person\\
                            Sister $\sqsubseteq$ Female\\
                            Sister $\sqsubseteq$ PersonWithASibling\\
                            Son $\sqsubseteq$ Child, Son $\sqsubseteq$ Male
                        \end{minipage}
                    };
                \end{tikzpicture}
            };
        \end{tikzpicture}
    \caption{A visualization of Family \ac{KB} with \TBox and a subset of \ABox. Colors denote concept assertions, while $(\cdot)$ and branching from $(\cdot)$ denote role assertions, respectively.}
     \label{fig:family_kb}
\end{figure}
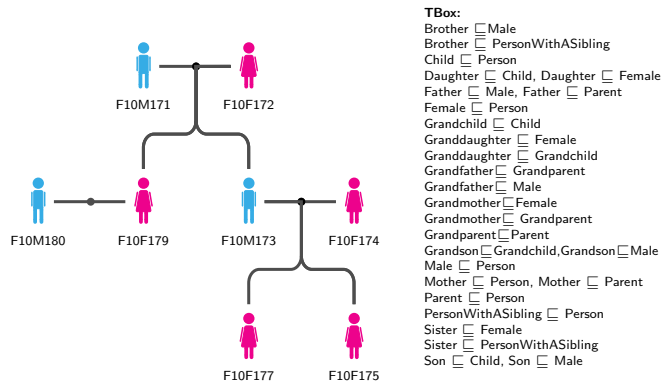

A Knowledge Base (\ac{KB}) is a pair $\kb=(\TBox,\ABox)$, where \TBox is a set of terminological axioms describing relations between named concepts $\NamedConcepts$~\cite{ijcai2021p281}.
A terminological axiom is in the form of $A\sqsubseteq B$ or $A\equiv B$ s.t.  $A,B \in \NamedConcepts$.
\ABox is a set of assertions describing relationships among individuals $a, b \in \Individuals$ via roles $r \in \Roles$ as well as concept membership relationships between $\Individuals$ and $\NamedConcepts$. 
Every assertion in \ABox must in the form of $A(x)$ and $r(x,y)$, where $A \in \NamedConcepts$, $r \in \Roles$, and $x,y \in \Individuals$.  
An example is visualized in~\Cref{fig:family_kb}.

\subsubsection{Description Logics:} Description Logics (\acp{DL}) are fragments of first-order predicate logic using only unary and binary predicates. 
The unary predicates, the binary predicates and constants are called concepts, roles and individuals, respectively~\cite{baader2003description}. 
\ac{DL} have become standard techniques to formalize background knowledge for many application domains including Semantic Web \cite{horrocks2003shiq,michel2019efficient}.
Leveraging \acp{KB} defined over \acp{DL} has a potential of being a backbone for explainable \ac{AI}~\cite{ijcai2021p281}.
Here, we consider \acp{KB} in the  
\ac{DL} \ALC (Attributive Language with Complements)~\cite{baader2003description} as in many other works (see~\Cref{sec:relatedwork}).
The model-theoretic semantics of \ALC are given in~\Cref{tab:semantics}.
\begin{table}[h]
	\centering
	\small
	\caption{$\mathcal{ALC}$ syntax and semantics. $\mathcal{I}$ stands for an interpretation, $\Delta^\mathcal{I}$ for its domain.}
	\label{tab:semantics}
   \begin{tabular}{l  c  c }
        \toprule
		Construct               & Syntax         & Semantics \\
		\midrule
        Atomic concept          & $A$            & $A^{\mathcal{I}}\subseteq{\Delta^\mathcal{I}}$\\
        Role                    & $r$            & $r^\mathcal{I}\subseteq{\Delta^\mathcal{I}\times \Delta^\mathcal{I}}$\\
 		Top concept             & $\top$         & $\Delta^\mathcal{I}$\\
 		Bottom concept          & $\bot$         & $\emptyset$            \\
 		Conjunction             & $C\sqcap D$    & $C^\mathcal{I}\cap D^\mathcal{I}$\\
 		Disjunction             & $C\sqcup D$    & $C^\mathcal{I}\cup D^\mathcal{I}$\\
 		Negation                & $\neg C$       & $\Delta^\mathcal{I}\setminus C^\mathcal{I}$\\
 		Existential restriction & $\exists~ r.C$ & $\{ x \mid \exists~y. (x,y) \in r^\mathcal{I} \text{ and } y \in C^\mathcal{I}\}$\\
 		Universal restriction & $\forall~ r.C$   & $\{ x \mid \forall~y. (x,y) \in r^\mathcal{I} \text{ implies } y \in C^\mathcal{I}\}$\\ 		
		\bottomrule
	\end{tabular}
\end{table}
\subsubsection{Concept Learning:} Let $\kb$ over \ALC, the set $E^+ \subset \Individuals$ of positive examples, and the set $E^- \subset \Individuals$ of negative examples be given.
The \ac{DL} concept learning problem is defined as follows
\begin{equation}
  \forall p \in E^+, \forall n \in E^- \big(\kb \models \texttt{H}(p)) \wedge  (\kb \not \models \texttt{H}(n) \big),
  \label{eq:cel}
\end{equation}
where
$\texttt{H} \in \AllALCConcepts$ denotes an \ALC concept and 
$\AllALCConcepts$ denotes all valid \ALC concepts under the
construction rules: $\texttt{C} :: = \texttt{A} \mid \neg \texttt{C} \mid \texttt{C} \sqcap \texttt{C} \mid \texttt{C} \sqcup \texttt{C}\mid \exists r.\texttt{C} \mid  \forall r.\texttt{C} \mid$, where $\texttt{A} \in \NamedConcepts$ and $r \in \Roles$.
$\kb \models \texttt{H}(p)$ implies that an inference of the class membership $\texttt{H}(p)$ is a logical consequence of $\kb$.
Checking whether a \texttt{H} fulfills \Cref{eq:cel}
is performed by a retrieval function $\Retrievalfunc : \; \AllALCConcepts \to 2^{\Individuals}$ defined under \ac{OWA} or \ac{CWA}.
This non-trivial learning problem is often transformed into a search problem within a quasi-ordered \ALC concept space~$(\StateSpace,\preceq)$
~\cite{Buhmann:2018,fanizzi2019boosting,lehmann2010concept,Tran2017Parallel}.
Traversing in \StateSpace is commonly conducted via a top-down refinement operator defined as $\rho:\StateSpace \rightarrow 2^\StateSpace$ with
\begin{equation}
\label{eq:refinement}
     \forall \texttt{A} \in \StateSpace : \rho(\texttt{A}) \subseteq \{ \texttt{B} \in \StateSpace \;|\; \texttt{B} \preceq \texttt{A} \}.
\end{equation}
State-of-the-art \ac{CL} models begin their search towards a $\texttt{H}$, after a search tree is initialized with the most general \ac{DL} concept ($\top$) as a root node. 
This search tree is iteratively built by selecting a node containing a quasi-ordered \ac{DL} concept with the highest heuristic value and adding its qualifying refinements as its children into a search tree~\cite{lehmann2010concept}.
\subsubsection{Heuristics:} 
A heuristic function is the key to an efficient search in $\StateSpace$ towards a $\texttt{H}$~\cite{lehmann2011class}. 
The number of explored concepts and runtimes are used as proxy
for the efficiency.
Various heuristic functions have been investigated~\cite{lehmann2011class,westphal2021simulated}.
Most heuristic functions of state-of-the-art models can be considered as myopic functions favoring syntactically short and accurate concepts.
Hence, they are prone to stuck in a local optimum~\cite{westphal2021simulated}.
For instance, the heuristic function of CELOE is defined as
\begin{equation}
\heuristic_\text{CELOE}(\texttt{A},\texttt{B}) = \text{Q}(\texttt{B}) + \lambda \cdot \big[ \text{Q}(\texttt{B})-\text{Q}(\texttt{A}) \big] - \beta \cdot |\texttt{B}|,
\label{eq:celoe}
\end{equation}
where $\texttt{A} \in \StateSpace$, $\texttt{B} \in \rho(\texttt{A})$.
$\beta > \lambda \geq 0$ and $\text{Q}(\cdot)$ denotes
a quality function (e.g. $F_1$ score or accuracy). 
Through $\text{Q}(\cdot)$ and $|\cdot|$, the search is steered based on solely \texttt{A} and \texttt{B} towards more accurate and syntactically shorter concepts. 
$F_1 (\cdot)$ is defined as 
\begin{equation}
F_1(\texttt{A})=\frac{\mid E^+ \cap \Retrievalfunc(\texttt{A}) \mid}{ \mid E^+ \cap \Retrievalfunc(\texttt{A}) \mid + 0.5 (\mid E^- \cap \Retrievalfunc(\texttt{A}) \mid + \mid E^+ \setminus \Retrievalfunc(\texttt{A}) \mid)}.
\label{eq:f1}
\end{equation}
As the size of \ac{KB} grows, runtimes of performing retrieval operations $\Retrievalfunc(\cdot)$ increase~\cite{bin2016towards,bin2017implementing,lehmann2010learning}.
Consequently, traversing in $\StateSpace$ becomes a computational bottleneck.
Therefore, reducing the number of explored concepts plays an important role to tackle to tackle \ac{CL} on \acp{KB}.
Although state-of-the-art models (e.g. CELOE) apply redundancy elimination and expression simplification rules to reduce the number of explored concepts, impractical long runtimes of state-of-the-art models still prohibit large-scale applications~\cite{hitzler2020neural}. 
Moreover, the selected assumption underlying \Retrievalfunc() also plays a role to tackle \ac{CL} on large \acp{KB}.
Due to the incomplete nature of \acp{KB},
\ac{OWA} seems to be a more suitable assumption~\cite{rudolph2011foundations}.
Yet, Using \ac{OWA} often makes membership queries computationally more challenging~\cite{fanizzi2008dl,lehmann2011class}.
Consequently, \ac{CWA} is often adopted in many recent works~\cite{Heindorf2021EvoLearner,kouagou2021learning,Tran2017Parallel}.
\section{Related Work}
\label{sec:relatedwork}
A plethora of works have investigated learning \acp{DL} concepts from a \ac{KB} and input examples. 
We refer to~\cite{baader2003description,hitzler2009foundations,krotzsch2012description} for an introduction.  
Most symbolic systems differ in the usage of heuristic functions and the design of the refinement operators~\cite{BadeaShan:2000,Buhmann:2018,fanizzi2008dl,fanizzi2018dlfoil,lehmann2011class,lehmann2010concept,iannone2007algorithm,Tran2017Parallel}. 
DL-Learner~\cite{lehmann2009dl} is regarded as the most mature and recent system for \ac{CL}~\cite{sarker2019efficient}. 
DL-Learner consists of several state-of-the-art models, including ELTL, OCEL, and CELOE. 
ELTL is based on a refinement operator for the \ac{DL} $\mathcal{EL}$ and uses a heuristic function that favors syntactically short concepts. 
CELOE builds on OCEL and ELTL and it applies a more sophisticated heuristic function.
CELOE is currently the best \ac{CL} model available within DL-Learner and often outperforms many state-of-the-art models including OCEL and ELTL in terms of the quality of learned expression, number explored concepts, and runtimes~\cite{westphal2021simulated,lehmann2016dl}.
The aforementioned approaches apply redundancy elimination and expression simplification rules to reduce the number of explored concepts.
Although applying redundancy elimination and expression simplification rules often reduce the number of explored concepts, these operations introduces more computation and long runtimes still prohibit large-scale applications~\cite{hitzler2020neural}. 
Most recent works have focused on treating the impractical runtimes in \ac{CL}. CLIP~\cite{kouagou2021learning} is a neural approach that serves as an addition to refinement-based approaches and supports pruning the search space by predicting the length of a possible goal state. 
EvoLearner~\cite{Heindorf2021EvoLearner} represents
a concept as an abstract syntax tree corresponding an individual of an evolutionary algorithm.
The initial population of individuals is obtained via biased random walks originating from $E^+$.
Westphal et al.~\cite{westphal2021simulated} design a Simulated Annealing based meta-heuristic to balance the exploration-exploitation trade-off during the search process.
In this work, we mainly evaluate \Nero against CELOE provided in DL-Learner for two reasons: (1) DL-Learner is regarded as the most mature and recent system for \ac{CL}~\cite{sarker2019efficient} and (2) most recently developed models are often evaluated w.r.t. the quality of concepts as well as runtimes. 
Yet, not reporting the number of explored concepts does not permit us to quantify whether a possible improvement through \Nero may stem from our novel idea or our efficient implementation. 
Consequently, in our experiments, we mainly compare \Nero against CELOE in terms of number of explored concepts, quality of learned concepts as well as runtimes.
\section{Methodology}
\label{sec:methodology}
\subsubsection{Motivation:} The goal in the \ac{CL} problem is to find a \ac{DL} concept $\texttt{H} \in \AllALCConcepts$ maximizing~\Cref{eq:f1}.
Here, we are interested in achieving this goal by learning permutation-invariant embeddings tailored towards predicting $F_1$ scores of pre-selected concepts.
Through exploring top-ranked concepts at first, we aim to find a goal concept can only with few retrieval operations.
If a goal state is not found within top-ranked concepts, the search tree of a state-of-the-art \ac{CL} model can be initialized with top-ranked concepts and $\top$ concept along with corresponding heuristic values.
By this, the standard search procedure can be started in more advantageous states, than the most general concept $\top$.
\subsubsection{Approach:}~\Cref{eq:f1} indicates that $F_1(\cdot)$ is invariant to the order of individuals in $E^+, E^-$, and $\Retrievalfunc(\cdot)$.
Previously, Zaheer et al.~\cite{zaheer2017deep} have proven that all functions being invariant to the order in inputs can be decomposed into
\begin{equation}
    f( \mathbf{x})=\phi \Big( \sum_{x \in \mathbf{x}} \psi(x) \Big),
    \label{eq:deepset}
\end{equation}
where $\mathbf{x}=\{ x_1, \dots, x_m\} \in 2^\mathcal{X}$ and $\phi(\cdot)$ and $\psi(\cdot)$ denote a set of input and two parameterized continuous functions, respectively.
A permutation-invariant neural network defined via 
\Cref{eq:deepset} still abides by the universal approximation theorem~\cite{zaheer2017deep}.
We conjecture that such neural network can learn permutation-invariant embeddings for sets of individuals (e.g. $E^+$ and $E^-$) tailored towards predicting $F_1$ scores of pre-selected concepts.
Through accurately predicting $F_1$ scores of pre-selected \ac{DL} concepts, possible goal concepts from pre-selected concepts can be detected without using $F_1(\cdot)$ and $\Retrievalfunc(\cdot)$.
With these considerations, we define \Nero as follows
\begin{align}
    \Nero(E^+, E^-) =& \sigma \bigg( \phi \Big( \sum_{x \in E^+} \psi(x) \Big) 
    -  \phi \Big( \sum_{x \in E^-}\psi(x) \Big) \bigg),
    \label{eq:approach}
\end{align}
where $\psi(\cdot): \Individuals \to \mathbb{R}^m$ 
and $\phi: \mathbb{R}^m \to [0,1]^{ \mid \targetexpressions \mid}$ denote an embedding look-up operation and an affine transformation, respectively.
$\targetexpressions$ represents the pre-selected \ac{DL} concepts.
The result of the translation operation denoted with $\mathbf{z} \in \mathbb{R}^m$ is normalized via the logistic sigmoid function $\sigma(\mathbf{z}) = \frac{1}{1 + \text{exp}(-\mathbf{z})}$. 
Hence,
$\Nero : 2^{\Individuals} \times 2^{\Individuals} \mapsto [0,1]^{\mid \targetexpressions \mid}$
can be seen as a mapping from two sets of individuals to $|\targetexpressions|$ unit intervals.
\Nero can be seen as a multi-task learning approach that leverages the similarity between multi-tasks, where a task in our case corresponds to accurately predicting the $F_1$ score of a pre-selected \ac{DL} concept~\cite{caruana1998multitask}.

The importance of learning representations tailored towards \emph{related} tasks 
has been well investigated~\cite{goller1996learning,caruana1998multitask}.
Motivated by this, we elucidate the process of selecting \ac{DL} concepts in ~\Cref{alg:taskcreation}.
We select such concepts that their canonical interpretations do not fully overlap (see the 4.th line).
As shown therein, \Nero can be trained on knowledge base defined over any \acp{DL} provided that $\Retrievalfunc(\cdot)$ and $\rho(\cdot)$ are given.
\begin{algorithm}
\small
\textbf{Input}: $\Retrievalfunc(\cdot), \, \rho(\cdot),\, d$, maxlength 
\textbf{Output}: $\targetexpressions$
\begin{algorithmic}[1] 
\STATE $\targetexpressions:= \{\texttt{C} \mid \texttt{C} \in \rho(\top) \wedge |\texttt{C}| \leq \text{ maxlength } \wedge 0<|\Retrievalfunc(\texttt{C})| \}$
\FOR{ \textbf{each} $ \texttt{A} \in \targetexpressions$}
    \FOR{ \textbf{each} $ \texttt{B} \in \targetexpressions$}
        \IF {$\Retrievalfunc(\texttt{A}) \not = \Retrievalfunc(\texttt{B})$}
            \FOR{ \textbf{each} $ \texttt{X} \in \{\texttt{A} \sqcap \texttt{B}, \texttt{A} \sqcup \texttt{B} \} $}
            \IF {$| \Retrievalfunc(\texttt{X})|> 0 \wedge \Retrievalfunc(\texttt{X}) \not \in \{\Retrievalfunc(\texttt{E}) \mid \texttt{E} \in \targetexpressions\} $}
                \STATE Add $\texttt{X}$ to $\targetexpressions$.
            \ENDIF
            \IF {$|\targetexpressions|=d$}
                \STATE \textbf{return} \targetexpressions
            \ENDIF
            \ENDFOR
        \ENDIF
\ENDFOR
\ENDFOR
\IF { $| \targetexpressions| < d$}
\STATE Go to the step (2).
\ENDIF
\end{algorithmic}
\caption{Constructing target \ac{DL} concepts}
\label{alg:taskcreation}
\end{algorithm}

\subsubsection{Training Process:}
Let $\mathcal{D} = \{ (E^+ _i,E^- _i,\mathbf{y}_i)\}_{i=1} ^N$ represent a training dataset, where a data point $(E^+,E^-,\mathbf{y})$ is obtained in four consecutive steps: 
\begin{inparaenum}[(i)]
        \item Sample $\texttt{C}$ from $\targetexpressions$ uniformly at random,
        \item Sample k individuals $E^+ \subset \Retrievalfunc(\texttt{C})$ uniformly at random,
        \item Sample k individuals $E^-\subset \Individuals \setminus E^+$ uniformly at random, and
        \item Compute $F_1$ scores $\mathbf{y}$ via
        \Cref{eq:f1} w.r.t. $E^+, E^-$, for $\targetexpressions$.
\end{inparaenum}
For a given $(E^+,E^-,\mathbf{y})$ and predictions $\hat{\mathbf{y}}:= \Nero(E^+, E^-)$, an incurred binary cross entropy loss.
Important to note that after training process, 
permutation-invariant embeddings of any \ALC \ac{DL} concepts can be readily obtained omitting the translation operation in \Nero, e.g. 
embeddings of
a \ac{DL} concept (e.g. $\text{Male}\sqcap\exists\text{hasSibling.Female}$) can be obtained via $\phi \big( \sum_{x \in \Retrievalfunc(\text{Male}\sqcap\exists\text{hasSibling.Female})} \psi(x) \big)$.
In our project page, we provided a 2D visualization of learned embeddings for the Family \ac{KB}.

\section{Experiments}
\label{sec:Experiments}
We based our experimental setup on~\cite{buhmann2016dl,lehmann2011class,Buhmann:2018} and used learning problems provided therein. 
An overview of the datasets is provided in~\Cref{table:datasets}.
To perform extensive comparisons between models, additional learning problems are generated by randomly sampling $E^+$ and $E^-$. 
We ensured that none of the learning problems used in our evaluation has been used in the unsupervised training phase. 
In our experiments, we evaluated all models in $\mathcal{ALC}$ for \ac{CEL} on the same hardware. 
\begin{table}[h]
\centering
\caption{An overview of class expression learning benchmark datasets.}
    \begin{tabular}{lcccc}
        \toprule
        Dataset & $|\Individuals|$ & $|\NamedConcepts|$ & $| \Roles |$  \\
        \midrule
        Family & 202 & 18 & 4\\
        Carcinogenesis & 22372 & 142 & 21\\
        Mutagenesis & 14145 & 86 & 11\\
        Biopax  & 323 & 28 & 49\\
        Lymphography&148&49&1 \\
        \bottomrule
    \end{tabular}
\label{table:datasets}
\end{table}

We evaluated models via the $F_1$ score, the runtime and number of explored concepts. 
The $F_1$ score is used to measure the quality of the concepts found w.r.t. positive and negative examples, while the runtime and the number of explored concepts are to measure the efficiency. We measured the full computation time including the time spent prepossessing time of the input data and tackling the learning problem. 
Moreover, we used two standard stopping criteria for state-of-the-art models. 
\begin{inparaenum}[(i)]
\item We set the maximum runtime to 10 seconds although models often reach good solutions within 1.5 seconds~\cite{lehmann2010concept}. 
\item The models are configured to terminate as soon as they found a goal concept.
\end{inparaenum}
In our experiments, we evaluate all models in $\mathcal{ALC}$ for \ac{CL} on the same hardware.
During training, we set $|\targetexpressions|=1000$, $N=50$ and used Adam optimizer for \Nero.
We only considered top-100 ranked concepts to evaluate \Nero.
\section{Results}
\subsubsection{Results with Benchmark Learning Problems:}\Cref{table:benchmark_results} reports the concept learning results with benchmark learning problems.
\Cref{table:benchmark_results} suggests that equipping \Nero with the standard search procedure improves the state-of-the-art performance in terms of $F_1$ scores even further with a small cost of runtimes.
CELOE and ELTL require at least $\mathbf{14.7 \times}$ more time than \Nero to find accurate concepts on Family.
This stems from the fact that \Nero explores on average only 21 concepts, whereas CELOE explored 1429.
On Mutagenesis and Carcinogenesis,~\Nero finds more accurate concepts, while exploring less, hence, achieving better runtime performance.
Runtime gains stem from the fact that \Nero explores at least $\mathbf{2.3 \times}$ fewer concepts. 
\begin{table}
    \caption{Results on benchmark learning problems. $F_1$, T, and Exp. denote $F_1$ score, total runtime in seconds, and the number of explored concepts, respectively. $\Nero^\dagger$ denotes equipping \Nero with CELOE. ELTL does not report the Exp.}
    \centering
    \begin{tabular}{lccccccccccccccccccccc}
    \toprule 
     \textbf{Dataset} & \multicolumn{3}{c}{$\textbf{\Nero}^\dagger$}&& \multicolumn{3}{c}{\textbf{\Nero}}&& \multicolumn{3}{c}{\textbf{CELOE}} && \multicolumn{2}{c}{\textbf{ELTL}}\\ 
    \cmidrule{2-4}  \cmidrule{6-8} \cmidrule{10-12} \cmidrule{14-15}
                     &$F_1$           & T & Exp. && $F_1$& T& Exp.&& $F_1$          & T     & Exp.    && $F_1$          & T     \\     
    \textbf{Family}  &$\mathbf{.987}$&.83   &26   &&.984  &$\mathbf{.68}$   &$\mathbf{21}$   &&$.980$         &$4.65$ &$1429$   &&$.964$         &$4.12$  \\     
    \textbf{Mutagenesis}     &$\mathbf{.714}$  &17.30   &200   &&.704  &$\mathbf{13.18}$  &$\mathbf{100}$  &&.704&23.05  & 516     &&.704& 21.04          \\     
    \textbf{Carcinogenesis}  &$\mathbf{.725}$  &32.23   &200   &&$.720$  &$\mathbf{26.26}$  &$\mathbf{100}$  &&.714           &37.18  & 230     && .719          & 36.29          \\     
    \bottomrule
    \end{tabular}
    \label{table:benchmark_results}
\end{table}

Important to note we did not use parallelism in \Nero and we reload parameters of \Nero for each single learning problem.
To conduct more extensive evaluation, we generated total 750 random learning problems on five benchmark datasets. 
Since Lymphography and Biopax datasets do not contain any learning problems, they are not included in~\Cref{table:benchmark_results}.
\subsubsection{Results with Random Learning Problems:}
\Cref{table:random_lp} reports the concept learning results with random learning problems.
\Cref{table:random_lp} suggests that CELOE explores at least $\mathbf{3.19 \times}$ more concepts than \Nero.
Importantly, \Nero finds on-par or more accurate concepts, while exploring less.
Here, we load the parameters of \Nero only once per dataset and are used to tackle learning problems sequentially.
This resulted in reducing the total computation time of \Nero by $\mathbf{3-6\times}$ on Family, Mutagenesis and Carcinogenesis benchmark datasets. 
Although \Nero can tackle learning problems in parallel (e.g. through multiprocessing), we did not use any parallelism, since CELOE and ELTL do not abide by parallelism~\cite{Buhmann:2018}.
Loading the learning problems in a standard mini-batch fashion and using multi-GPUs may further improve the runtimes of \Nero.
These results suggest that \Nero can be more suitable than CELOE and ELTL on applications requiring low latency.
\begin{table}
    \caption{Random learning problems with different sizes per benchmark dataset. 
    Each row reports the mean and standard deviations attained in 50 learning problems.
    $|E|$ denotes $|E^+| + |E^-|$. 
    }
    \centering
    \resizebox{\columnwidth}{!}{
    \begin{tabular}{llccccccccccccccccccccccc}
    \toprule 
     \textbf{Dataset} &$|E|$&\multicolumn{3}{c}{\textbf{\Nero}}&& \multicolumn{3}{c}{\textbf{CELOE}} && \multicolumn{2}{c}{\textbf{ELTL}}\\ 
        \cmidrule{3-5} \cmidrule{7-9} \cmidrule{11-12}
                        &   &$F_1$                     & T                    & Exp.                && $F_1$           & T             & Exp.         && $F_1$           & T           \\
\textbf{Family}         &10 &$\mathbf{.913\pm.06}$&$\mathbf{.16\pm.51}$&$\mathbf{ 74\pm43}$  &&$.903\pm.06$&$11.61\pm3.58$ &$5581\pm2375$ &&$.718\pm.01$&$4.45\pm2.84$\\
                        &20 &$\mathbf{.807\pm.04}$&$\mathbf{.16\pm.49}$&$\mathbf{100\pm00}$   &&$.795\pm.05$&$13.28\pm1.47$ &$7586\pm645$  &&$.678\pm.02$&$3.59\pm1.27$\\
                        &30 &$\mathbf{.775\pm.03}$&$\mathbf{.15\pm.41}$&$\mathbf{100\pm00}$   &&$.760\pm.03$&$13.24\pm1.42$ &$7671\pm575$  &&$.672\pm.01$&$3.46\pm1.59$\\
\midrule
\textbf{Lymphography}  &10 &$\mathbf{.968\pm.07}$ &$\mathbf{.12\pm.43}$&$\mathbf{ 75\pm41}$  &&$.968\pm.07$&$ 6.63\pm4.29$ &$5546\pm5169$ &&$.733\pm.09$&$3.07\pm.30$\\
                      &20 &$\mathbf{.828\pm.04}$ &$\mathbf{.13\pm.40}$&$\mathbf{100\pm00}$  &&$.826\pm.05$&$13.01\pm1.23$ &$11910\pm1813$&&$.678\pm.02$&$3.08\pm.50$\\ 
                      &30 &$\mathbf{.780\pm.04}$ &$\mathbf{.13\pm.01}$&$\mathbf{100\pm00}$  &&$.780\pm.04$&$13.02\pm1.69$ &$13138\pm2601$&&$.672\pm.01$&$3.09\pm.72$\\
\midrule
\textbf{Biopax}        &10 &$\mathbf{.859\pm.08}$ &$\mathbf{.19\pm.71}$&$\mathbf{ 86\pm34} $ &&$.806\pm.07$&$13.26\pm1.94$ &$4752\pm2153$ &&$.685\pm.06$&$3.71\pm .10$\\         
                      &20 &$\mathbf{.793\pm.05}$ &$\mathbf{.19\pm.52}$&$\mathbf{100\pm00}$  &&$.746\pm.04$&$13.63\pm.10$ &$4151\pm748$  &&$.668\pm.06$&$3.72\pm .10$\\
                      &30 &$\mathbf{.749\pm.03}$ &$\mathbf{.18\pm.52}$&$\mathbf{100\pm00}$  &&$.718\pm.02$&$13.91\pm.44$ &$3843\pm963$  &&$.668\pm.06$&$3.90\pm .22$\\
\midrule
\textbf{Mutagenesis}   &10 &$\mathbf{777\pm.05}$ &$\mathbf{3.47\pm1.61}$&$\mathbf{100\pm00}$  &&$.753\pm.06$&$20.27\pm1.39$ &$546\pm613$   &&$.670\pm.02$&$10.29\pm.40$\\
                      &20 &$\mathbf{.746\pm.05}$ &$\mathbf{3.09\pm1.75}$&$\mathbf{100\pm00}$  &&$.712\pm.02$&$20.38\pm1.30$ &$430\pm 28$   &&$.667\pm.00$&$10.73\pm1.10$\\
                      &30 &$\mathbf{.721\pm.03}$ &$\mathbf{2.89\pm1.60}$&$\mathbf{100\pm00}$  &&$.700\pm.02$&$20.39\pm1.06$ &$429\pm 38$   &&$.667\pm.00$&$11.74\pm.97$\\
\midrule
\textbf{Carcinogenesis}&10 &$\mathbf{.768\pm.06}$ &$\mathbf{5.39\pm2.98}$&$\mathbf{98\pm14}$  &&$.764\pm.06$&$29.90\pm1.02$ &$401\pm125$   &&$.673\pm.05$&$19.99\pm.67$\\
                      &20 &$\mathbf{.722\pm.03}$&$\mathbf{5.40\pm1.87}$&$\mathbf{100\pm00}$  &&$.713\pm.02$&$30.30\pm.19$ &$318\pm152$   &&$.667\pm.00$&$20.00\pm1.11$ \\
                      &30 &$\mathbf{.704\pm.05}$ &$\mathbf{4.70\pm2.78}$&$\mathbf{100\pm00}$  &&$.697\pm.02$&$29.99\pm.58$ &$319\pm 43$   &&$.667\pm.00$&$20.38\pm.85$ \\
\bottomrule
\end{tabular}}
\label{table:random_lp}
\end{table}
\begin{table}
    \caption{Performance comparison with different number of explored concepts.
    Each row reports the mean and standard deviations attained in 50 learning problems.}
    \centering
    \resizebox{\columnwidth}{!}{
    \begin{tabular}{llccccccccccccccccccccccc}
    \toprule 
     \textbf{Dataset} &$|E|$&\multicolumn{2}{c}{\textbf{\Nero-1}}&& \multicolumn{2}{c}{\textbf{\Nero-10}} && \multicolumn{2}{c}{\textbf{\Nero-1000}}\\ 
        \cmidrule{3-4} \cmidrule{6-7} \cmidrule{9-10}
                      &     &$F_1$            & T            && $F_1$          & T          && $F_1$   & T   \\
\textbf{Family}       &10   &$.906\pm.07$  &$\mathbf{.08\pm.06}$    &&$.910\pm.05$ &$.09\pm.06$ &&$\mathbf{.916\pm.06}$&$.81\pm.50$\\
                      &20   &$.793\pm.05$  &$\mathbf{.08\pm.05}$    &&$.806\pm.04$ &$.09\pm.05$ &&$\mathbf{.807\pm.04}$&$1.17\pm.50$\\
                      &30   &$.742\pm.05$  &$\mathbf{.08\pm.05}$    &&$.773\pm.03$ &$.09\pm.05$ &&$\mathbf{.775\pm.03}$&$1.15\pm.50$\\
\midrule
\textbf{Lymphography} &10   &$.882\pm.07$  &$\mathbf{.08\pm.06}$    &&$.905\pm.05$ &$.08\pm.06$ &&$\mathbf{.916\pm.06}$&$.77\pm.50$\\
                      &20   &$.793\pm.05$  &$\mathbf{.07\pm.05}$    &&$.827\pm.04$ &$.08\pm.05$ &&$\mathbf{.828\pm.04}$&$1.03\pm.50$\\
                      &30   &$.738\pm.05$  &$\mathbf{.07\pm.06}$    &&$.777\pm.04$ &$.08\pm.05$ &&$\mathbf{.780\pm.03}$&$1.00\pm.60$\\
\midrule
\textbf{Biopax}       &10   &$.853\pm.08$  &$\mathbf{.09\pm.06}$    &&$.856\pm.05$ &$.97\pm.59$ &&$\mathbf{.868\pm.08}$&$1.31\pm.80$\\
                      &20   &$.779\pm.05$  &$\mathbf{.09\pm.06}$    &&$.791\pm.04$ &$.10\pm.62$ &&$\mathbf{.793\pm.04}$&$1.35\pm.60$\\
                      &30   &$.708\pm.07$  &$\mathbf{.09\pm.06}$    &&$.742\pm.04$ &$.10\pm.63$ &&$\mathbf{.749\pm.03}$&$1.39\pm.60$\\
\midrule
\textbf{Mutagenesis} &10    &$.733\pm.07$  &$\mathbf{.32\pm2.03}$    &&$.785\pm.06$  &$.57\pm1.81$ &&$\mathbf{.803\pm.06}$  &$36.98\pm5.81$\\
                     &20    &$.689\pm.08$  &$\mathbf{.31\pm1.99}$    &&$.734\pm.05$  &$.52\pm1.82$ &&$\mathbf{.751\pm.04}$  &$34.29\pm5.91$\\
                     &30    &$.673\pm.08$  &$\mathbf{.31\pm2.03}$    &&$.712\pm.04$  &$.49\pm1.77$ &&$\mathbf{.728\pm.03}$  &$32.88\pm6.13$\\
\midrule
\textbf{Carcinogenesis}&10  &$.717\pm.09$  &$\mathbf{.41\pm2.53}$&&$.740\pm.09$&$.89\pm2.89$&&$\mathbf{.783\pm.02}$&$56.941\pm9.55$\\
                      &20   &$.680\pm.06$  &$\mathbf{.40\pm2.49}$&&$.707\pm.05$&$.82\pm2.45$&&$\mathbf{.731\pm.02}$&$57.205\pm5.08$\\
                      &30   &$.610\pm.11$  &$\mathbf{.41\pm2.83}$&&$.671\pm.06$&$.77\pm2.66$&&$\mathbf{.716\pm.02}$&$52.872\pm7.58$\\
\bottomrule
\end{tabular}}
\label{table:random_lp_diff_exp}
\end{table}
\subsubsection{Results with Limited Exploration:}
\Cref{table:random_lp_diff_exp} reports concept learning results with limited exploration on five benchmark datasets.
~\Cref{table:random_lp_diff_exp} suggests that \Nero-10
often outperforms CELOE and ELTL (see~\Cref{table:random_lp}) in all metrics even when exploring solely 10 top-ranked concepts.  
\subsubsection{Significance Testing:}
To validate the significance of our results, we performed Wilcoxon signed-rank tests (one and two-sided) on $F_1$ scores, runtimes and the number of explored concepts.
Our null hypothesis was that the performances of \Nero and CELOE come from the same distribution. 
We were able to reject the null hypothesis with a p-value $< 1\%$ across all the datasets, hence, the superior performance of \Nero is statistically significant.
\subsection{Discussion} 
Our results uphold our hypothesis: 
$F_1$ scores of \ac{DL} concepts can be accurately predicted by means of learning permutation-invariant embeddings for sets of individuals.
Through considering top-ranked \ac{DL} concepts at first, the need of excessive number of retrieval operations to find a goal concept can be mitigated.
Throughout our experiments,~\Nero consistently outperforms state-of-the-art models w.r.t. the $F_1$ score, the number of explored concepts and the total computational time.
Importantly, starting the standard search procedure on these top-ranked concepts further improves the results.
Hence, \Nero can be applied within state-of-the-art models to decrease their runtimes.
However, it is important to note that Lehmann et al.~\cite{lehmann2011class} have previously proved the completeness of CELOE in the \ac{CL} problem, i.e., for a given learning problem, CELOE finds a goal expression if it exists provided that there are no upper-bounds on the time and memory requirements. 
Although these requirements are simply not practical, equipping \Nero with the search procedure of CELOE is necessary to achieve the completeness in \ac{CL}.
\section{Conclusion}
We introduced a permutation-invariant neural embedding model (\Nero) to efficiently tackle the description logic concept learning problem.
For given learning problem, \Nero accurately predicts $F_1$ scores of pre-selected description logic concepts in a multi-label classification fashion.
Through ranking concepts in descending order of predicted $F_1$ scores, a goal concept can be learned within few retrieval operations.
Our experiments showed that~\Nero outperforms state-of the art models in 770 concept learning problems on 5 benchmark datasets w.r.t. the quality of predictions, number of explored concepts and the total computational time. 
Equipping \Nero with the standard search procedure further improves the $F_1$ scores across learning problems and benchmark datasets.

We believe that incorporating neural models in concept learning problems is worth pursuing further. 
In future, we will work on using \Nero on more expressive description logics
and integrating embeddings for concepts in non-myopic heuristics~\cite{demir2021drill}.
\section*{Acknowledgments}
This work has been supported by the European Union’s Horizon Europe research and innovation programme (GA No 101070305), by the Ministry of Culture and Science of North Rhine-Westphalia within the project SAIL (GA No NW21-059D), and the Deutsche Forschungsgemeinschaft (GA No TRR 318/1 2021 – 438445824).
\bibliographystyle{splncs04}
\bibliography{references}
\end{document}